\documentclass[prb,twocolumn,showpacs,showkeys]{revtex4}%
\usepackage{graphicx}
\usepackage{amsmath}
\usepackage{amsfonts}
\usepackage{amssymb}

\date{\today}

\begin{document}
\title[Dipolar interaction effects in magnetic nanoparticle arrays]
{Dipolar interaction effects in the magnetic and magnetotransport properties of ordered nanoparticle arrays}

\author{D. Kechrakos}
\email{dkehrakos@ims.demokritos.gr}
\author{K. N. Trohidou}
\affiliation{Institute of Materials Science NCSR "Demokritos", GR-153 10, Athens, Greece}

\keywords{magnetic nanoparticles; ordered arrays; self-assembly; anisotropy; exchange bias; dipolar interactions;  transverse susceptibility; tunneling magnetoresistance; resistor network; Monte Carlo}
\pacs{75.50.Tt, 75.75.+a, 75.20.-g, 75.47.-m}

\begin{abstract}

Assemblies of magnetic nanoparticles exhibit interesting physical properties arising from the competition of intraparticle dynamics and interparticle interactions.
In ordered arrays of magnetic nanoparticles  magnetostatic interparticle interactions introduce collective dynamics acting competitively to random anisotropy. Basic understanding, characterization and control of dipolar interaction effects in arrays of magnetic nanoparticles is an issue of central importance. To this end, numerical simulation techniques offer an indispensable tool. We report on Monte Carlo studies of the magnetic hysteresis and spin-dependent transport in thin films formed by ordered arrays of magnetic nanoparticles. Emphasis is given to the modifications of the single-particle behavior  due to interparticle dipolar interactions as these arise in quantities of experimental interest, such as, the magnetization, the susceptibility and the magnetoresistance. We investigate the role of the structural parameters of an array (interparticle separation, number of stacked monolayers) and the role of the internal structure of the nanoparticles (single phase, core-shell). Dipolar interactions are responsible for anisotropic magnetic behavior between the in-plane and out-of-plane directions of the sample, which is reflected on the investigated magnetic properties (magnetization, transverse susceptibility and magnetoresistance) and the parameters of the array (remanent magnetization, coercive field, and blocking temperature). Our numerical results are compared to existing measurements on self-assembled arrays of Fe-based and Co nanoparticles is made.
\end{abstract}
\maketitle

\section{INTRODUCTION}

Laterally confined magnetic nanostructures (dots and nanoparticles) is an important class of novel materials with unique physical properties, that emerge because their size becomes comparable to various characteristic physical lengths (correlation length, domain wall width, etc). Owing to their novel physical properties they find numerous technological applications in magnetic storage media\cite{THO00}, magnetic sensors\cite{KIR00} and magnetic logic devices.\cite{COW06}

Magnetic nanoparticles (NPs) are commonly formed in assemblies, with either random or ordered structure. In the first group belong systems such as ferrofluids and granular solids, while in the second group belong the patterned media (or magnetic dots) and the self-assembled arrays (SAA) of NPs. The existence of order in a NP assembly is a decisive property in view of their application in magnetic storage media with ultrahigh density $(\sim Tb/in^2)$, which rely on the possibility of treating the NPs as individually addressable magnetic bits.

Magnetic dots are developed by lithographic processes and they are characterized by lateral dimensions in the range of $10-100~nm$.\cite{MAR03} The major advantage of this approach to development of ordered nanostructured materials is that it offers great flexibility and good control over the shape, the size and the arrangement of the dots as well as the choice of the constituent material. However, shape imperfections and formation of polycrystalline dots are the major factors that determine the deviations from perfect periodicity in the arrays. Furthermore, the size of the magnetic dots, determined by the limitations of the lithographic process is in most cases comparable to the exchange length, thus permitting the domain formation in the ground state or during magnetization reversal, a fact that makes the study of the magnetic behavior of dot arrays quite intricate. Growth methods and magnetic properties of isolated dots and dot arrays have been recently reviewed by Martin \emph{et al}.\cite{MAR03}

In magnetic NPs with diameter $D\approx 1-10~nm$, the domain wall width is well beyond the diameter of the NP, and consequently domain formation at the ground state or during magnetization reversal is precluded. The magnetization of the NPs is practically at its saturation value up to temperatures less than but close to the bulk Curie temperature. They are commonly referred to as single-domain (SD) particles, to distinguish them from magnetic dots that exhibit inhomogeneous magnetization.
Highly monodisperse ($\sigma \lesssim 5\%$) SD magnetic NPs are produced by solution chemistry methods and organized in hexagonal ordered arrays by self-assembly.\cite{PET98,SUN99,HAR01,PUN01a}
The small size, the high monodispersity and the array periodicity achieved by self-assembly, as well as the low production cost of SAA, has motivated a great deal of research effort in the field of synthesis and magnetic characterization of these systems.
The synthetic routes and the structural and magnetic characterization methods have been recently reviewed by Willard \emph{et al}\cite{WIL04}.

Assemblies of magnetic nanoparticles have been investigated in an effort to gain basic understanding of the interplay between single-particle magnetic anisotropy and interparticle magnetostatic interactions. To this end, a variety of sample preparation methods have been adopted by research groups in order to grow and magnetically characterize nanoparticle assemblies, as for example, frozen ferrofluids \cite{DOR97,BAT02,JON04}, discontinuous metal-insulator multilayers\cite{KLE01,CHE03,SAH03,PTR06}, co-sputtered metal-insulator films\cite{SAN00,LUI02,LUI02b}, cluster-assembled films\cite{BAN05,BIN05} self-organized particle arrays on surfaces\cite{HAU98} and chemically produced self-assembled nanoparticle arrays.\cite{PET98,PIL01,SUN99,PUN04,FAR05, MAJ06} Investigations of the static and dynamic magnetic properties of dipolar interacting nanoparticle assemblies brought up fundamental issues related to the existence of a ground state which shares common features with canonical spin glasses (slow relaxation, memory and ageing effects). \cite{KLE01,BIN02,CHE03,SAH03,JON04}. Dipolar interparticle interactions are considered responsible for the observed complex (spin-glass-like) behavior of sufficiently dense and random nanoparticle assemblies. In view of the technological importance of the hysteresis behavior and the thermal stability of magnetization of NP assemblies\cite{MOS02}, the important issue of the effects of magnetostatic interactions  on the static magnetic properties has also been studied extensively.\cite{DOR97,BAT02}

Contrary to random assemblies, ordered arrays of NPs are ideal systems to investigate the role of interparticle interactions, for two reasons. First, the NP arrangement is periodic with small perturbations, and the undesired complications introduced by spatial randomness are substantially suppressed. Second, chemically synthesized magnetic NPs are often coated by an inorganic surfactant layer that prevents agglomeration during self-assembling but also keeps the surfaces of neighboring particles at a distance well beyond the range of exchange forces. As a consequence, the prevailing interparticle interactions in a SAA are magnetostatic. Finally, the spherical, in most cases, shape of the NPs diminishes the importance of higher order multipolar interactions and the assembly is well described by  dipolar interparticle forces.

Various experiments have demonstrated the presence of magnetostatic interactions in SAA with various degrees of structural disorder and layered NP assemblies. Reduction of the remanence at low temperature,\cite{HEL01} increase of the blocking temperature,\cite{MUR01,LUI02,ZHA03} increase of the barrier distribution width,\cite{WOO01} deviations of the zero-field cooled magnetization curves from the Curie behavior,\cite{PUN01a,PUN01b} difference between the in-plane and out-of-plane remanence,\cite{RUS00} and increase of the blocking temperature with frequency of applied field\cite{POD02} have been observed and attributed to interparticle dipole-dipole interactions (DDI). Long-range ferromagnetic order in linear chains\cite{HAU98,COW00,RUS03} and hexagonal arrays\cite{RUS00,PUN04,FAR05,MAJ06} of dipolar coupled single-domain magnetic nanoparticles has been demonstrated, supporting the existence of a dipolar superferromagnetic ground state.
Nanoparticle assemblies with random morphology have been studied more as most growth techniques developed so far (sputtering, cluster beams, mechanical alloying) produce random samples. Ordered nanoparticle arrays, on the other hand, have been less studied both theoretically and experimentally due to the difficulty in producing ordered samples. Chemical synthesis and self-assembly offer a new and promising approach to this direction.\cite{BAD05} We therefore believe that basic understanding of the magnetic properties of dense (interacting) ordered arrays is currently highly demanded.

In the ongoing research effort for development of magnetic nanostructures with reduced size and improved thermal stability\cite{MOS02}, the exploitation of the exchange bias effect in laterally confined structures (dots and nanoparticles) has attracted a lot of interest.\cite{NOG05}
Atomic scale models of the magnetic structure have been developed in an effort to interpret  experimental observations of the exchange bias effect in composite NPs with a ferromagnetic (FM) core and an antiferromagnetic (AFM) shell.
Among the most important theoretical results\cite{EFT05} we mention
(i) the disappearance of the exchange bias field $(H_E)$ at temperatures above the N\'{e}el temperature of the AFM, in agreement with experiments\cite{LIE03},
(ii) the strong dependence of $H_E$ on the number unsaturated bonds across the FM-AFM interface and the dependence of $H_C$ on the interface area,
(iii) the increase of both $H_E$ and $H_C$ for a given NP radius with increasing oxidation depth, (iv) the increase of $H_E$ and decrease of $H_C$ with increasing oxidation layer thickness and a fixed core radius,
(v) the fast stabilization of $H_E$ with increasing core size, in agreement with experiments\cite{MOR04}, and
(vi) the reduction of $H_C$ and increase of $H_E$ and its thermal stability with increasing exchange constant of the AFM material and/or at the FM-AFM interface.
Despite the research effort focused on the atomic scale mechanism of magnetization reversal in composite nanoparticles,\cite{ZIA98,TRO02,EFT05,IGL05} much less attention has been paid so far to the modification of the magnetic hysteresis behavior due to inter-particle interactions arising in assemblies. In this direction, Fe NPs embedded in iron-oxide matrix\cite{BIA02} were shown to freeze below a temperature owing to the competition between the exchange anisotropy at the core-shell interface and the interparticle DDI. Similarly, increase of the exchange bias field due to magnetostatic interparticle coupling was found in stripes of Co/CoO nanoparticles\cite{HBI03} and inter-dot magnetostatic interactions were shown to produce asymmetric anomalies in the magnetization reversal mechanism of Co/CoO dot arrays.\cite{GIR03}
The modification of the coercive and exchange-bias fields in dense nanoparticle arrays with core-shell morphology as a result of the competition between exchange anisotropy and interparticle dipolar interactions consists a challenging issue.

Detection and quantification of DDI in assemblies of magnetic NPs has been addressed so far by a variety of experimental techniques including in most cases SQUID magnetometry and AC susceptibility measurements\cite{DOR97}, and more recently small-angle neutron scattering (SANS)\cite{IJI05} and resonant magnetic X-ray scattering\cite{KOR05}. The last method is a direct probes of magnetic correlations at the interparticle scale. These studies have provided ample evidence that the interplay between random anisotropy and DDI determine the magnetic behavior of the NP assemblies.
More recently dipolar interaction effects and the resulting collective dynamics in SAA of Fe \cite{POD03}, Fe$_2$O$_3$ \cite{SPI01}, and Co \cite{SPI02a,SPI02b} NPs was studied by reversible transverse susceptibility (RTS) measurements. The RTS technique is a well established and powerful method to obtain information about the anisotropy of magnetic nanoparticles, from considerations of the peak positions of the field-dependent RTS.\cite{AHA57} The implementation of RTS to study SAA arrays revealed information regarding the different dynamical regimes of an interacting assembly accessed as the temperature increases.

Electron spin is a degree of freedom whose control and detection in transport measurements is the basis of the rapidly developing field of spintronics\cite{WOL01}. Charge transport measurements in a SAA of Co NPs were performed\cite{BLA00} and revealed a spin-dependent tunneling mechanism which is responsible for substantial $(\sim 10\%)$ tunneling magnetoresistance (TMR) values at low temperature $(\sim 20K)$. The tunneling barriers are provided in an array by the insulating surfactant layer surrounding the NPs. TMR measurements probe the interparticle correlations within the range of the spin-diffusion length, and are therefore sensitive to the magnetic microstructure of the assembly. Interparticle interaction effects are expected to reveal themselves in the TMR signal. The field-dependent magnetization and conductivity were discussed in the experiments of Black \emph{et al}\cite{BLA00}, however a systematic correlation between the two quantities remains to be performed.

In addition to the experimental work, various numerical studies that focused on the ground state configuration and the hysteresis behavior of dipolar interacting nanoparticle arrays have appeared. The interplay of DDI and perpendicular anisotropy was shown\cite{STA99} to induce a reorientation transition below a critical temperature and interaction-induced shape anisotropy of a finite sample controls the magnetization reversal mode. Dipolar interactions were found to decrease the coercive field of magnetic nanoparticle arrays independently of the array topology (square or hexagonal) despite the fact that the ground state configuration is determined by the array topology.\cite{RUS01} The presence of an incomplete second layer with hexagonal structure does not destroy the long-range FM ordering of the ground state,\cite{KEC02} while even slight structural disorder within the array destroys that ordering.\cite{JEN03} On the other hand, higher order (quadropolar) magnetostatic interactions were shown to act in synergy with DDI stabilizing the long range order of the ground state in a nanoparticle array.\cite{POL02}
Previous theoretical studies of RTS in random assemblies of magnetic NPs demonstrated that a wide size distribution rounds the peaks of RTS\cite{HOA93}, orientational texture suppresses the coercivity peak \cite{HOA93} and dipolar interactions lead to merge of the coercivity and anisotropy peaks\cite{CHA94,YAN95}. More recently the issue of the structure of the RTS curves of SAA was addressed by Monte Carlo (MC) simulations \cite{KEC06} that reproduced many of the experimental observations from RTS measurements in SAA of Fe NPs.\cite{POD03}.
Charge transport in nanoparticle arrays has been studied by resistor network models (RN) that include in a phenomenological way the essential aspects of the thermally-activated hopping mechanism\cite{HEL76}, the spin-dependence of the hopping proccess\cite{INO96} and the details of the micromagnetic configuration of the sample.\cite{KEC05} In a recent study\cite{KEC05} the signature of dipolar interaction effects in TMR measurements has been investigated.

In this article we review our results from MC simulations of the field and temperature dependence of the magnetization, the RTS and the TMR of hexagonal arrays of dipolar coupled magnetic nanoparticles with random anisotropy. The consideration of a hexagonal arrangement of NP is an essential feature of our model as DDI have a well known anisotropic character that relates their magnitude and sign to the relative position of the interacting dipoles. The main structural parameters we focus on are (a) the interparticle separation, which is directly related to the dipolar coupling strength and can be experimentally controlled by variation of the surfactant layer during the synthetic process, and (b) the sample thickness, namely the number of stacked monolayers (MLs), which is a crucial parameter for the collective response of the array, controlled by the NP concentration in the colloidal dispersion.
The aim of the present work is to reveal the modification introduced to experimentally measured properties of an ordered NP assembly (magnetization, susceptibility, magnetoresistance) due to the presence of DDI. The remaining of the paper is organized as follows: In Section II we present the structural and magnetic model used in our simulations. In Section III we discuss numerical results on the hysteresis characteristics (saturation remanence, coercivity), the zero field cooled (ZFC) and field cooled (FC) curves and and the extracted blocking temperature $(T_b)$. The magnetization of the interesting class of composite nanoparticles with a core-shell morphology is also discussed. The evolution of the RTS curves with temperature and dipolar strength is discusses next, and finally magnetoresistance calculations are presented. Whenever experimental results are available they are compared with our simulations aiming to reveal the character of interparticle interactions in different measured samples. Final conclusions and remarks are given in the Section IV.

\section{MODEL AND SIMULATION METHOD}

We proceed with the definition of the spin model used to describe the magnetic structure of a magnetic NP array formed by either simple ferromagnetic or composite (FM core/ AFM shell) nanoparticles.
The NPs forming an ordered array are assumed spherical and monodisperse. The size dispersion is not expected to introduce major modifications to the magnetic behavior because particularly low values are achieved $(\sigma \lesssim 5\%)$ in most samples.\cite{SUN99,WIL04,FAR05} The NP diameter is $D$ and they occupy the sites of a triangular lattice in the $xy$-plane with lattice constant $d \geq D$. When more that one MLs are considered, the particles are close-packed in an $ABCABC...$ stacking sequence. This structure is consistent with electron microscopy studies of Co\cite{MUR01,PUN01a,PUN01b} and Fe\cite{YAM02} NP arrays. Incomplete layer of NPs are formed by random occupancy of the triangular lattice sites.
The NPs are single domain with uniaxial anisotropy in a random direction, and they interact via dipolar forces. The total energy of the system is
\begin{eqnarray}
E = g\sum_{ij}\frac{ \widehat{S}_{i}  \cdot \widehat{S}_{j}
 - 3(\widehat{S}_{i} \cdot \widehat{R}_{ij} )
    (\widehat{S}_{i} \cdot \widehat{R}_{ij} )  } {R_{ij}^{3}} \nonumber \\
 - k \sum_{i} ( \widehat{S}_{i}  \cdot \widehat{e_{i}} )^{2}
 - h \sum_{i} ( \widehat{S}_{i}  \cdot \widehat{H} )
\label{eq1}
\end{eqnarray}
where $\widehat{S}_{i}$ is the magnetic moment direction (spin) of the $i$-th particle, $\widehat{e_{i}}$ is the easy-axis direction, and ${R}_{ij}$ is the center-to-center distance between particles $i$ and $j$. Hats in Eq~(1), and further on, indicate unit vectors. The energy parameters entering Eq.~(\ref{eq1}) are the dipolar energy $g=m^{2} /d^{3}$, where $m =M_{s}V$ is the particle moment, the anisotropy energy $k=K_{1}V$, and the Zeeman energy $h=mH$ due to the applied dc field $H$. The energy parameters ($g,k,h$) entering Eq.~(\ref{eq1}), the thermal energy $t=k_{B}T$, and the history of the sample determine the micromagnetic configuration at a certain temperature and bias field. Because, our simulation method relies on minimization of the free energy of the system, multiplication of all the energy parameters by the same scaling factor does not modify the results. Thus, in all subsequent results we scale the energy parameters entering Eq.~(\ref{eq1}) by the single particle anisotropy energy $(k=1)$. This choice makes our numerical results applicable to a class of materials with the same parameter ratios rather than to a specific material.
The crucial parameter that determines the transition from single-particle to collective behavior is the ratio of the dipolar to the anisotropy energy $g/k=(\pi/6)(M_{s}^{2}/K_{1})(D/d)^{3}$.
The reported values \cite{RUS00,PUN01a,ZHA03} for fcc or hcp Co NPs are $g/k= 0.2-0.4(D/d)^{3}$, while for the soft $\epsilon$-Co phase, higher values are expected.\cite{PUN01a} For Fe NPs Farrell \emph{et al} \cite{FAR05} report $g/k= 1.54~(D/d)^{3}$.
Despite the relative dispersion of the reported values, the important issue is that for most samples of Co and Fe NPs the ratio of dipolar to anisotropy strengths is below unity, except for $\epsilon$-Co. Thus, in the numerical results presented in the following section we consider $g/k$ values less than one.

Extensions to the spin model described by Eq.~(\ref{eq1}) are required in order to study composite NPs with a FM core and an AFM shell.
In the present work we adopt a model introduced by Meiklejohn and Bean\cite{MEI57} (further on referred to as the MB model) in their interpretation of shifted loops observed in oxidized transition metal NPs samples after zero-field cooling. The MB model provides a \emph{phenomenological} understanding of the exchange bias effect and the unidirectional anisotropy.\cite{NOG99}  Consequently, important parameters of the exchange-bias effect such as the interface structure and interface magnetization are averaged out. Despite its simplicity, the MB model and its variations was successfully implemented in the case of FM/AFM bilayers to interpret the dependence of the exchange bias field on temperature \cite{ZAA96}, on the thickness the AFM layer\cite{BNK01} and on the direction of the applied field. \cite{HUJ02}. The major weaknesses of the MB model being the overestimation of the exchange-bias field values,\cite{NOG99} and the underestimation of the coercivity values.\cite{COMMENT-1} We adopt the MB model as the simplest possible approach to bring out the essential aspects of the competition between intraparticle (uniaxial anisotropy) and interparticle (dipolar) interactions. Our purpose, is to investigate this interplay, which is expected to be important in dense samples, rather than revealing the atomic scale mechanism which is responsible for the exchange bias effect.

According to the MB model, coherent rotation of the atomic spins is assumed in the FM core and the AFM shell, while the net magnetic moment of the shell is vanishingly small. In addition, the interface of the AFM is assumed fully  uncompensated, namely all spins belong to the same sublattice and is exchanged coupled to the core. Consequently, the magnetic state of each NP is described by a pair of anisotropic and exchange coupled spins, $S^{FM}$ and $S^{AFM}$. For a dipolar interacting assembly the total energy reads
\begin{eqnarray}
E = g\sum_{ij}\frac{ \widehat{S}_{i}^{FM}  \cdot \widehat{S}_{j}^{FM}
 - 3(\widehat{S}_{i}^{FM} \cdot \widehat{R}_{ij} )
    (\widehat{S}_{j}^{FM} \cdot \widehat{R}_{ij} )  } {R_{ij}^{3}}
\nonumber \\
 - k_{C} \sum_{i} ( \widehat{S}_{i}^{FM}  \cdot \widehat{e_{i}} )^{2}
 - k_{S} \sum_{i} ( \widehat{S}_{i}^{AFM} \cdot \widehat{e_{i}} )^{2}
\nonumber \\
 - J \sum_{ij}   \widehat{S}_{i}^{FM} \cdot  \widehat{S}_{i}^{AFM}
 - h \sum_{i}   \widehat{S}_{i}^{FM}  \cdot \widehat{H}
\label{eq2}
\end{eqnarray}
where $\widehat{S}_{i}^{FM}$ and $\widehat{S}_{i}^{AFM}$ indicate the magnetization directions (spins) of the core and the interface layer of the shell, respectively. $J$ is the interface exchange energy and $k_C, k_S$ are the distinct values of the core and shell anisotropy, respectively. Notice that owing to the zero net magnetization of the AFM shell, only the FM cores couple to the external field and between them via magnetostatic forces.
For simplicity we have assumed in Eq.~(\ref{eq2} that the core and shell magnetizations have a common easy axis, that is therefore labeled by the particle index $i$. Due to the vanishing net magnetization of the shell there is neither Zeeman nor dipolar contributions to the total energy due to the shell.  However, the coupling across the FM-AFM interface makes an exchange contribution to the total energy, expressed by the fourth term in Eq.~(\ref{eq2}).

For either simple or composite NP assemblies, the magnetic configuration is obtained by a MC simulation, using the standard Metropolis algorithm.\cite{MMC88} The initial spin configuration corresponds to the saturation state along a chosen axis ($x$ or $z$).
Experimentally, to observe a shifted loop a field-cooling process is performed prior to the hysteresis measurement in order to align the AFM moments parallel to the moments of the FM. \cite{NOG99,NOG05} Furthermore, the value of the observed exchange bias field i.e. the loop shift) increases with the value of the cooling field $(H_{FC})$.\cite{NOG99,NOG05}  The choice of the saturation state as the initial state to calculate the hysteresis loop of core-shell NPs from the fully saturated state, it is equivalent to assuming an infinitely strong cooling field ($H_{FC} \gg J_{AFM}$). Thus, the maximum value of $H_E$ is obtained.
During relaxation, the initial $10^{3}$ MC steps per spin (MCS) are used for relaxation of the system towards equilibrium and thermal averages are calculated over the subsequent $10^{4}$ MCS, allowing 10 MCS between sampling events to achieve statistical independence. The results are averaged over $N_{c}=30-100$ samples with different realizations of the random axes distribution and the thermal fluctuations. To deal with the long-range character of the DDI we use periodic boundaries in the $xy$-plane and implemented the Ewald summation method adapted to a quasi-two-dimensional system.\cite{GAO97,GRZ00,LOM00} Free boundaries along the $z$-axis are assumed.

\begin{figure}
\includegraphics[]{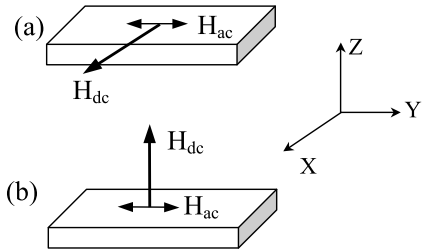}
\caption{Sketch of the sample and applied magnetic field used in our simulations. (a) In-plane and (b) out-of-plane directions of the bias  field ($H_{dc}$). The indicated ($H_{ac}$) field is considered only in the calculation of the transverse susceptibility. For the magnetoresistance calculations a dc bias voltage is applied on opposite edges of the sample along the $y$-axis.}
\label{f1}
\end{figure}

Transverse susceptibility measurements are performed with an ac measuring field $(H_{ac})$ perpendicular to the dc bias field (see, Fig.~\ref{f1}). The ac field is weak $(\sim10~Oe)$ and its frequency lies in the rf regime ($f \sim 10^6$ Hz).\cite{SPI00,SPI01}. The weak measuring field permits us to neglect transverse hysteresis effects and calculate the RTS in the zero-field limit. Furthermore, since that the N\'eel relaxation time of the NP magnetization is large compared to the inverse frequency of the ac field the static approximation for the measuring field is justified. Susceptibility values are obtained from the fluctuations of the magnetization $M_{y}=\sum S_{i}^{y}$, as \cite{YEO92}
\begin{eqnarray}
\chi_T^{\parallel (\perp)}(H_{x(z)})
=\frac{1}{N k_{B}T} [ \langle M_{y}^{2} \rangle - \langle M_{y} \rangle^{2} ]
\label{eq3}
\end{eqnarray}
where $\chi_T^{\parallel (\perp)}$ is the in-plane (out-of-plane) RTS and $N$ is the number of NPs in the simulation cell.

In the last part of this section we describe the resistor network model used to calculate the magnetoresistance of the array. For a given micromagnetic configuration $\{ \widehat{S}_i \}$ we introduce the spin-dependent conductivity between two nanoparticles $i$ and $j$ as\cite{INO96},
\begin{equation}
\sigma_{ij}=\sigma_0(1+P^2cos\theta_{ij})~exp(-R_{ij}/\alpha-E_c/k_BT)
\label{eq4}
\end{equation}
where
$\sigma_0=2e^2 /h$ is the conductivity quantum,
$P$ is the spin polarization of the conduction electrons,
$cos\theta_{ij}=(\hat{S}_i \cdot \hat{S}_j)$ ,
$E_c=e^2/2C$ is the the activation energy to charge a neutral NP by addition of a single electron,
$C$ is the NP capacitance relative to its surrounding medium, and
$\alpha=\hbar/\sqrt{8m^*(U-E_F)}$ is the electron wave function decay length in the insulating barrier of height $U$ relative to the Fermi energy.
In all our simulations we take $\alpha=d$, as a sufficient requirement to allow charge transfer between neighboring nanoparticles and $P=0.34$ which is an appropriate for Co NPs.\cite{MES73,MUR01}
Charge conservation on every node of the network implies
\begin{equation}
\sum_{ij}\sigma_{ij}(\phi_i-\phi_j)=0
\label{eq5}
\end{equation}
where $\phi_i$ is the local value of the electric potential. Eq.~(\ref{eq5}) is solved for the local potentials with the boundary conditions that set the local potential on opposite sides of the sample, namely at $y=0$ and $y=L$, (see Fig.~\ref{f1} equal to zero and $\phi_0$, respectively. The total conductivity is given as
\begin{equation}
\sigma=\frac{1}{2\phi_0^2} \sum_{ij}\sigma_{ij}(\phi_i-\phi_j)^2=0.
\label{eq6}
\end{equation}
Obviously the values of $\sigma$ depend on the spin configuration {$\hat{S}_i$}. A thermal average of the conductivity is obtained by averaging the conductivity values, as obtained from Eq.~(\ref{eq6}) over a sequence of equilibrium spin configurations produced by the MC algorithm. Finally, the tunneling magnetoresistance of the sample is defined as
\begin{equation}
TMR(H)=\frac{\sigma_{sat} - \sigma(H)} {\sigma(H)}
\label{eq7}
\end{equation}
where $\sigma_{sat}$ is the saturation value of the conductivity. It follows from Eq.~(\ref{eq4}) that the local conductivity between particles $i$ and $j$ increases quadratically with spin polarization $(P)$ and exponentially with localization length $(\alpha)$  and particle capacitance $(C)$, the latter depending on the NP diameter.\cite{BLA00} The TMR values are expected to have a similar dependence on the parameters  $\alpha$ and $P$. However, for a monodisperse sample the TMR values are independent of $C$. Since we are mainly interested here is the shape of the field-dependent TMR curves, rather than the actual values of TMR, the dependence of the TMR on $\alpha$  and P is not considered further on.

\section{NUMERICAL RESULTS AND DISCUSSION}

\subsection{Isothermal hysteresis and ZFC/FC magnetization of FM nanoparticles}

\begin{figure}
\includegraphics[]{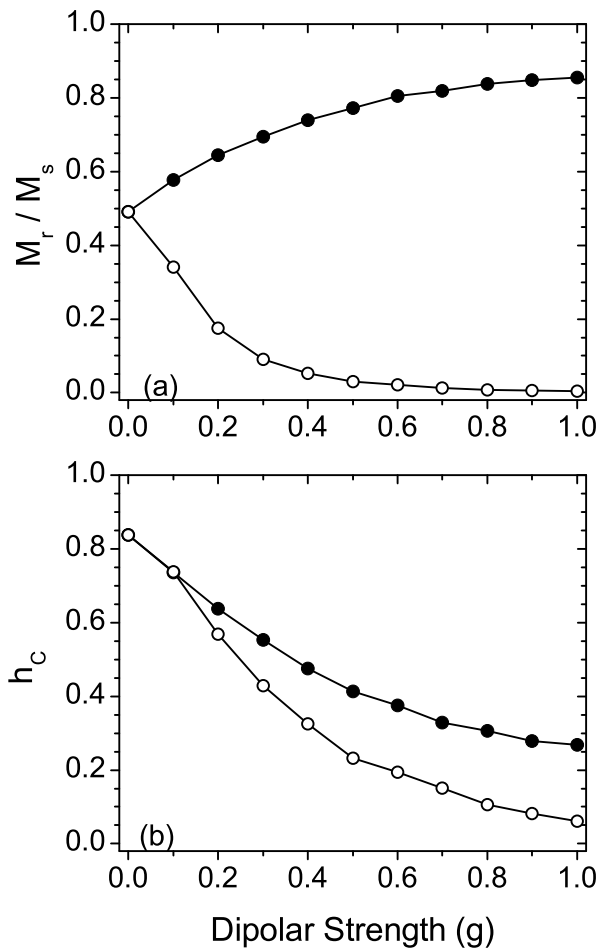}
\caption{Dependence of (a) saturation remanence and (b) coercivity on the dipolar strength at low temperature $(t/k=0.01)$ for a monolayer of FM nanoparticles.  Closed symbols : In-plane field. Open symbols : Out-of-plane field}
\label{f2}
\end{figure}

Two important characteristics of the hysteresis loop are the saturation remanence $(M_r)$ and the coercivity $(H_c)$. The former is the magnetization of the system after removal of a saturating field, and the latter the negative field required to zero the magnetization of a system being in the positive remanent state before application of the field.  In Fig.~\ref{f2} we show the dependence of $M_r$ and $H_c$  on the strength of the dipole-dipole interaction. Variation of the dipolar strength in a SAA can be achieved experimentally in two different ways, namely, by varying the thickness of the surfactant layer\cite{SUN00}, or by synthesis of different material NPs with the same size and the same surfactant layer thickness, or combination of both.
In Fig.~\ref{f2} we show that depending on the direction of the applied field these quantities show different variation with dipolar strength.
This behavior can be explained by the ferromagnetic and anisotropic character of the DDI. It is a well established \cite{ROZ91,RUS01,POL02} fact that DDI on a triangular lattice stabilize a FM ground state and create an easy-plane for the array magnetization due to their anisotropic character. Therefore, the in-plane remanence tends towards the saturation value $(M_r/M_s=1)$ as the dipolar strength increases, while the out-of-plane remanence decreases continuously because the out-of-plane field is normal to the easy-plane. On the other hand, the coercivity decreases with increasing dipolar strength, independently of the applied field direction. This behavior can be understood as due to collective rotation of the moments. For an in-plane field, the effective anisotropy of a cluster of dipolar coupled NPs is reduced (as a result of an averaging process over many random easy directions) and the total moment of a cluster is larger than a single NP. Due to the synergy of these two factors a weaker reversal field is required. For an out-of-plane field, the development of an easy-plane forces the moments to lie in the $xy$-plane reducing their projection along the field axis, thus a weak field is required to fully zero the magnetization. Anisotropy between the in-plane and out-of-plane remanence $(M_r^{\parallel} > M_r^{\perp})$ has been observed in arrays of Co NPs\cite{RUS00}, where a ratio of $\gamma \equiv M_r^{\perp} / M_r^{\parallel} = 0.30$ was found. For these samples the reported\cite{RUS00} dipolar strength is $g/k=0.07$ and from the data shown in Fig.~\ref{f2} we obtain $\gamma(g=0.07)=0.3$ in good agreement with the experiments. A zero-temperature calculation\cite{RUS00} on a triangular lattice gave a similar value for $\gamma$. In the same experiments\cite{RUS00} a negligibly small dependence of the coercivity on the applied field direction was found. Our simulations agree with this feature showing, that for dipolar strengths in the range $g/k=0 - 0.1$, $H_C^{\parallel}$ and $H_C^{\perp}$ practically coincide (Fig.~\ref{f1}). Our MC simulations and the energy minimization approach\cite{RUS00} justify the dominant role of DDI in the magnetic properties of these arrays.

\begin{figure}
\includegraphics[]{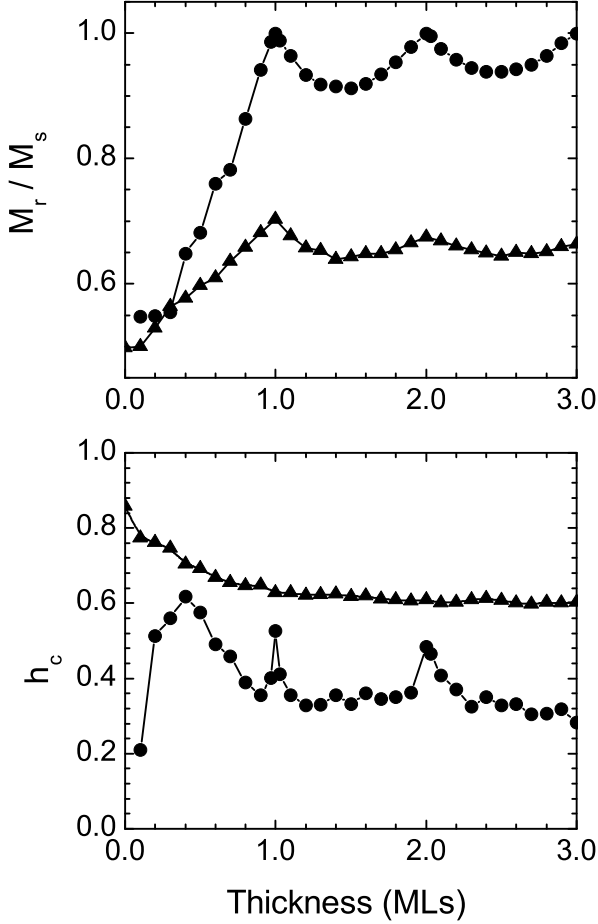}
\caption{Dependence of saturation remanence (upper panel) and coercivity (lower panel) on number of stacked monolayers of FM nanoparticles, at low temperature ($t/k=0.01$). Non-integer values of film thickness correspond to a randomly occupied uppermost layer. Triangles: weak coupling $(g/k=0.25)$. Circles: strong coupling $(g/k=10)$. The applied field lies in-plane. }
\label{f3}
\end{figure}

An important structural parameter in SAA is the number of stacked MLs.
This parameter can be partially controlled either by varying the solvent concentration\cite{PUN01a,PUN01b,PIL01,MUR01} prior to self-assembly or by forming the arrays using the Langmuir-Blodgett technique.\cite{FRI01,POD02} To address the effects of film thickness on the hysteresis properties of NP arrays we show in Fig.~\ref{f3} the variation of the remanence and coercivity with the number of MLs.
For strong dipolar coupling the remanence takes maximum values when complete monolayers form, while an incomplete top layer suppresses the magnetic order due to the competing character of the DDI in a random nanoparticle assembly\cite{KEC98}. Peaks in $H_c$ are observed at full coverage, which indicate that fully ordered samples are magnetically harder relative to samples with disordered surfaces.
The observed decay of $H_c$ with increasing layer thickness marks a transition from a two-dimensional reversal mode ($c = 1$ML), during which the moments are forced by the interaction-induced easy-plane anisotropy to remain in-plane during reversal, to a three-dimensional mode ($c \geq 2$ ML), during which the rotation path of the moments is not restricted in the plane of the film.
When DDI are weak, it is shown in Fig.~\ref{f3} that the oscillatory dependence of $M_r$ is suppressed and the peaks in  $H_c$ are washed out. However, the decay of $H_c$ with increasing thickness remains, reaching a constant value above $c\simeq 2$ MLs.
The increase of $M_r$ with coverage, observed for submonolayer coverage in Fig.~\ref{f3} is in agreement with experiments on dilute samples of Fe NPs\cite{FAR05} with variable concentrations. Also SAA of Fe NPs always showed higher $M_r$ values compared to dilute (disordered) samples.\cite{FAR05}. This trend is reproduced by our simulation results in Fig.~\ref{f3}.

\begin{figure}
\includegraphics[]{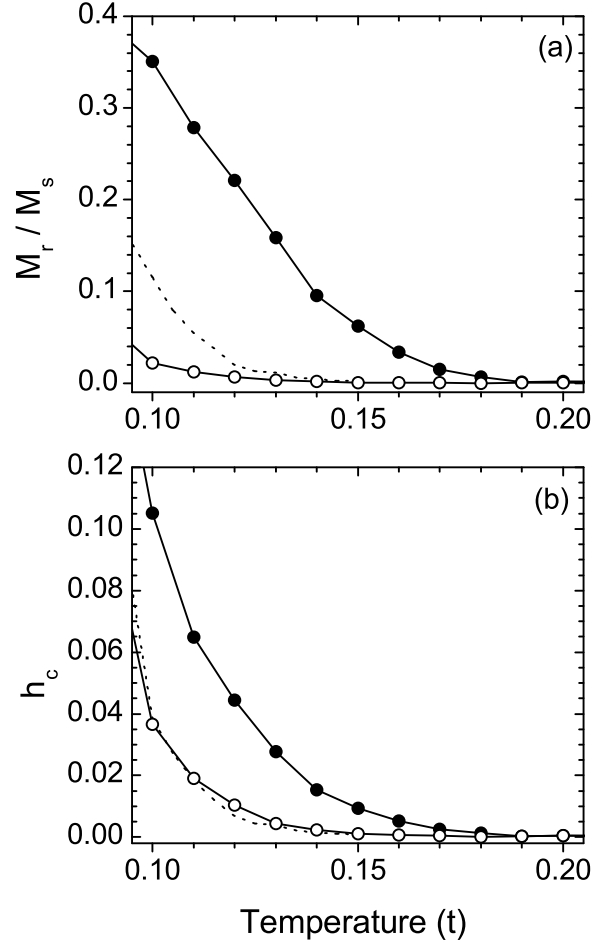}
\caption{Temperature dependence of (a) saturation remanence and (b) coercivity of a monolayer of weakly dipolar $(g=0.1)$ FM nanoparticles. Closed symbols: In-plane field. Open symbols: Out-of-plane field. Dashed line: Non-interacting nanoparticles.}
\label{f4}
\end{figure}

As temperature increases thermal fluctuations can assist the magnetic moments to overcome the anisotropy barrier leading the system to thermal equilibrium. The regime in which this is achieved is defined by the blocking temperature of the system. For dipolar interacting NPs the concept of a single-particle barrier becomes rather vague as the thermal activation of a moment is correlated to the motion of all the other moments of the system. However, one can still refer to the blocking temperature of the system in a phenomenological way, namely one can define it as the temperature above which the remanence and coercivity vanish.
We show in Fig.~\ref{f4} the temperature dependence of the in-plane and out-of-plane $M_r$ and $H_c$ for a monolayer of weakly coupled NPs. We first notice that the effect of interactions is to increase the blocking temperature of the array $(t_b\simeq0.17)$ relative to the non-interacting case $(t_b\simeq0.14)$. This is clear in Fig.~\ref{f4} for the in-plane field. Most interestingly, at temperatures above $t_b^0$, $M_r$ and $H_c$ of the interacting system are non zero.

This result defines an interesting temperature regime, $t_b^0 \leq t \leq t_b$, in which the thermal energy overwhelms the nominal anisotropy barrier $(E_b\simeq k)$, but the hysteresis behavior of the NP assembly persists due to DDI. Calculations of the energy barrier distribution in dipolar interacting NP assemblies showed that, DDI broaden the distribution thus producing a range of high barriers $(E_b > k)$, which are responsible for the observed enhanced thermal stability in dense assemblies.\cite{CHU05} Notice finally, that the enhanced thermal stability is less pronounced, but also existent in the out-of-plane geometry, despite the fact that in this geometry dipole interactions have a demagnetizing effect at low temperature (Fig.~\ref{f2}). A similar enhancement of $M_r$ and $H_c$ at elevated temperatures was previously predicted also for NP assemblies with random morphology\cite{KEC98} and its existence was verified experimentally from the measured increase of the blocking temperature with NP density in frozen ferrofluids.\cite{DOR97}

\begin{figure}
\includegraphics[]{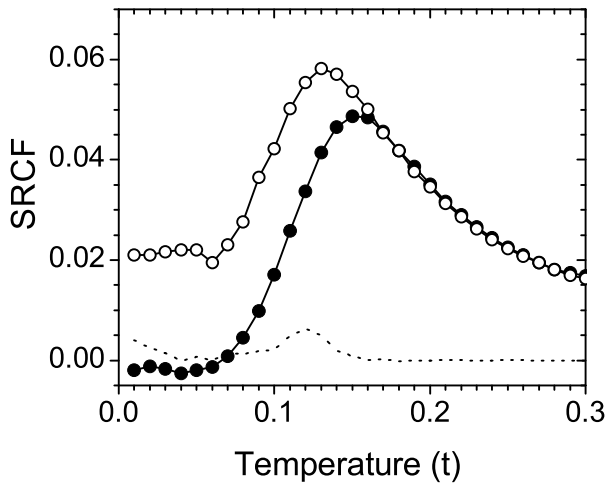}
\caption{Temperature dependence of the first nearest-neighbor correlation function at the remanent state for a monolayer of weakly dipolar $(g=0.1)$ FM nanoparticles. Closed symbols: In-plane field. Open symbols: Out-of-plane field. Dashed line: non-interacting assembly.  }
\label{f5}
\end{figure}

Further insight into the collective behavior at elevated temperatures can be extracted from examination of the short-range moment correlation function, defined as
\begin{equation}
SRCF = \langle \widehat{S}_{i} \cdot  \widehat{S}_{j} \rangle_{R_{ij}=d}
-\langle \widehat{S}_{i}  \rangle^{2}.
\label{eq8}
\end{equation}

Strictly speaking, the blocking temperature of a system corresponds to the maximum value of the long range moment correlation function, namely the susceptibility. We prefer to examine the short-range correlation function because it is directly accessible by various spectroscopic experiments to be mentioned below. The position of the  $SRCF$ peak of gives a reasonable approximation  to the blocking temperature of the system. The temperature dependence of the $SRCF$, shown in Fig.~\ref{f5}, shows that short range FM correlations exist above the blocking temperature of the non-interacting array $(t_b^0 \simeq 0.14)$ and persist up to temperatures above the blocking temperature of the interacting array $(t_b^{\parallel} \simeq 0.17)$. Recently, Kortright \emph{et al}\cite{KOR05} extracted the interparticle magnetic correlations in dense arrays of Co NPs from x-ray scattering experiments. They concluded that for $\epsilon$-Co NPs, which are strongly dipolar, AFM correlations exist at temperatures above the blocking.
SANS studies of Fe NP assemblies showed no evidence of AFM correlations at elevated temperature.\cite{FAR06}
Finally, our simulations indicate the existence of short-range FM interparticle correlations in dipolar coupled arrays.
There seems to be difficult at present to reach a decisive conclusion about the nature of interparticle correlations in NP arrays from experimental findings.
Scattering experiments (x-rays, SANS) at lower Q-values, are expected to probe magnetic correlations at a scale lying well within the inter-particle separation regime are most probably required in order to compare with our present simulations

An interesting feature shown in Fig.~\ref{f5} is the different temperatures at which the peaks of $SRCF$ are located. The relative shift of the peaks indicates that the array of dipolar interacting NPs exhibits anisotropic blocking behavior with $T_b^{\parallel} > T_b^{\perp}$.

\begin{figure}
\includegraphics[]{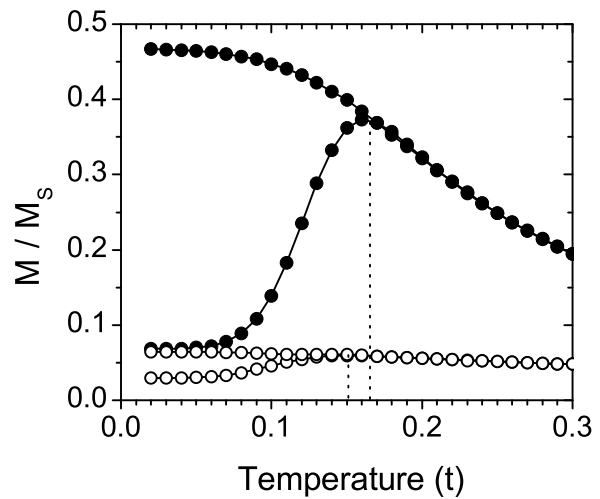}
\caption{ZFC/FC magnetization (at $h/k=0.1$) for a monolayer of weakly interacting $(g=0.1)$ FM nanoparticles.  Closed symbols: In-plane field. Open symbols: Out-of-plane field. Dashed lines indicate the different values of the in-plane and out-of-plane blocking temperatures. }
\label{f6}
\end{figure}

The anisotropic blocking can also be demonstrated by examination of the ZFC/FC curves shown in Fig.~\ref{f6}. The existence of anisotropic blocking for dipolar interacting arrays of Fe NPs has been recently demonstrated experimentally
in dense arrays of Fe NPs.\cite{POD03} In these experiments a ratio $T_b^{\parallel} / T_b^{\perp} \simeq 1.15$  was obtained from ZFC/FC measurements. The Fe NP parameters given\cite{POD03} are $M_s=1360~emu/cc$, $K=3.4\cdot 10^4~erg/cc$, $D=6.8~nm$ and $d\simeq20~nm$, which lead to $g/k \simeq 0.11$. Our simulations shown in Fig.~\ref{f6} for $g/k=0.1$ give $t_b^{\parallel} / t_b^{\perp} \simeq 1.08$, which is in reasonable agreement with the experiments, given various factors not considered in our model, such as the sample thickness, the deviations from perfect stacking of the monolayers and the in-plane structural defects.

\begin{figure}
\includegraphics[]{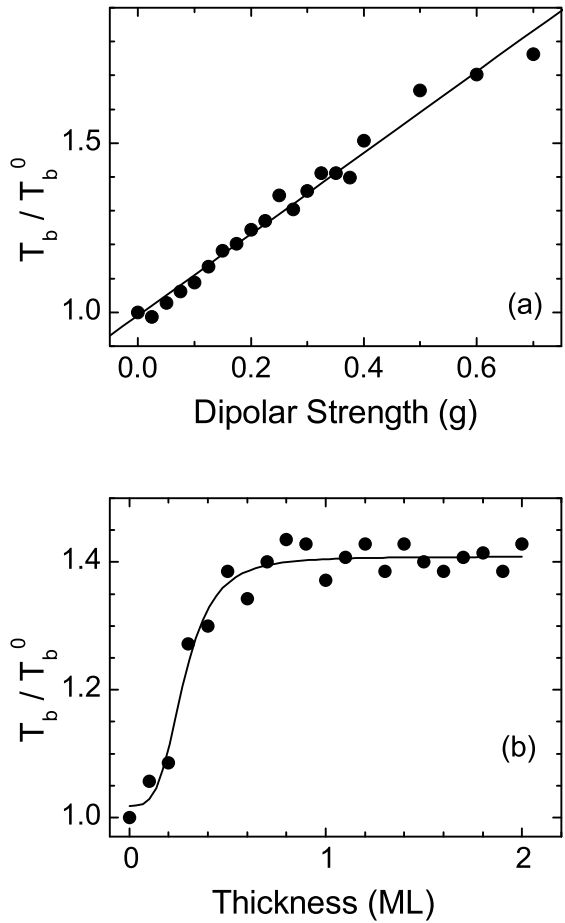}
\caption{(a) Dependence of blocking temperature (a) on dipolar strength, for a monolayer of FM nanoparticles, and (b) on film thickness, for FM nanoparticles with $g/k=0.25$. Measuring in-plane field $h/k=0.1$. The values of $T_b$ are normalized to the value corresponding to non-interacting nanoparticles.  }
\label{f7}
\end{figure}

Changing the structural parameters of a self-assembled array, namely interparticle spacing and film thickness is expected to affect the blocking behavior.
The blocking temperature is proportional to the effective barrier height which is determined by the single-particle anisotropy and the local dipolar field.
Dipolar interactions on the other hand are sensitive to interparticle spacing $(g \sim 1/d^3)$.
Decreasing the interparticle spacing is equivalent to increasing the dipolar strength,and consequently modifying the blocking temperature.
Similarly, increasing the film thickness we shown above that it suppresses at low-temperature the values of $M_r$ and $H_c$.
In Fig.~\ref{f7} we show the effect the structural changes have on the blocking temperature of the assembly, the latter being obtained from the peak of the corresponding ZFC curve.
We see that in the range of dipolar strengths considered here, $T_b$ scales linearly with the dipolar strength, or equivalently, it decreases with the cube of the interparticle separation $(T_b \sim 1/d^3)$.
We can therefore think of the effect of weak dipole interactions $(g/k <1)$ as an increase of the single-particle anisotropy barrier by an amount proportional to their coupling strength.

A similar effect on $T_b$ to decreasing the interparticle separation is obtained by increasing the areal coverage and the film thickness.
As shown in Fig.~\ref{f7} a dramatic increase of $T_b$ is observed during formation of the first monolayer and a saturation behavior is reached as soon as the second complete monolayer is formed.
Reduction of $T_b$ upon dilution of chemically synthesized assemblies has been reported by several groups\cite{SUN99,MUR01,ZHA03,FAR05}.
In particular, Zhang \emph{et al}\cite{ZHA03} performed ZFC/FC measurements on self-assembled $\epsilon$-Co NP and report $30\%$ increase of $T_b$ relative to highly dilute samples.
The NP parameters for these samples\cite{ZHA03} correspond to $g/k\sim0.2$, thus the predictions of our simulations (Fig.~\ref{f7}) are in agreement with these experiments.
Similar dependence of $T_b$ on NP concentration has been also observed for NPs dispersed in a solid matrix \cite{DOR97} and reproduced by simulations for 3D disordered assemblies of magnetic NPs.\cite{KEC98,CHA00}.
It is interesting that despite the demagnetizing character of the dipolar interaction in the ground state of 3D random assemblies\cite{KEC98,ELH99}, they tend to stabilize the FM character of the assembly at elevated temperatures $(T < T_b^0)$.
The increase of $T_b$ with number of stacked monolayers was observed in discontinuous Co-Al$_2$O$_3$ multilayers\cite{LUI02}. However, the slow saturation of $T_b$ obtained after 5-7 monolayers in these experiments is probably due to deviations from the ideal stacking sequence and the in-plane randomness in NP size and location, inherent to the sample preparation technique. We also believe that the lack of perfect stacking is probably the reason that measurements of $T_b$ in Langmuir-Blodgett films of Co NPs showed a decrease of $T_b$ with increasing thickness.\cite{POD02}

\subsection{Isothermal hysteresis of FM/AFM nanoparticles}

\begin{figure}
\includegraphics[]{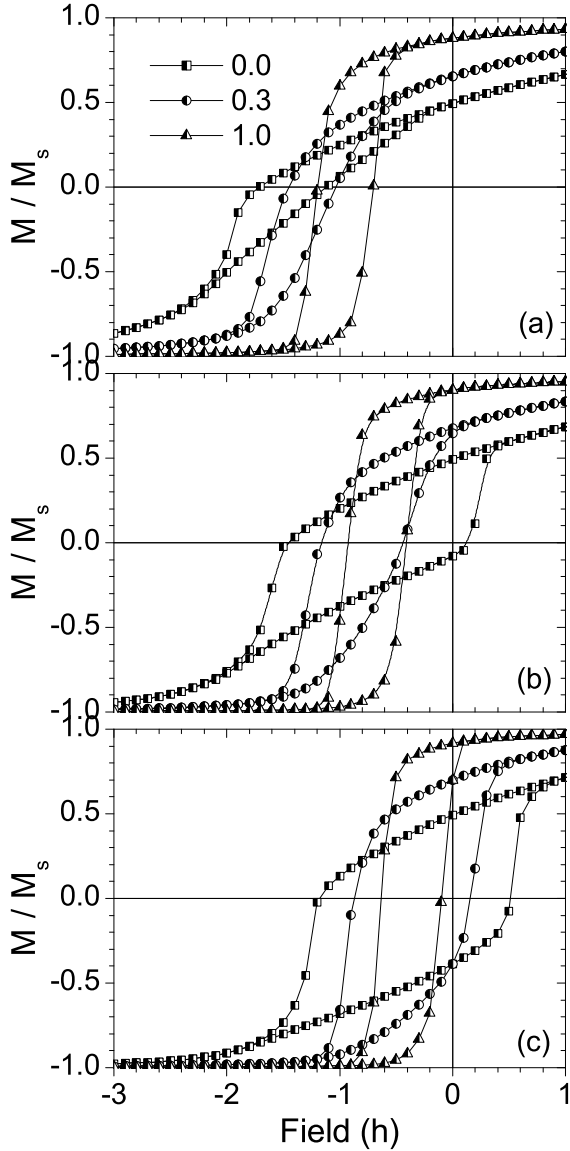}
\caption{Hysteresis loops at low temperature ($t/k_C=0.01$) of a monolayer of core-shell NPs with different dipolar strengths and interface exchange values. (a) $J/k_C=1.5$ , (b) $J/k_C=1.0$ and (c) $J/k_C=0.5$. Squares: $g=0$, circles: $g/k_C=0.3$, and triangles: $g/k_C=1.0$. Shell anisotropy $k_S/k_C=5.0$. }
\label{f8}
\end{figure}

\begin{figure}
\includegraphics[]{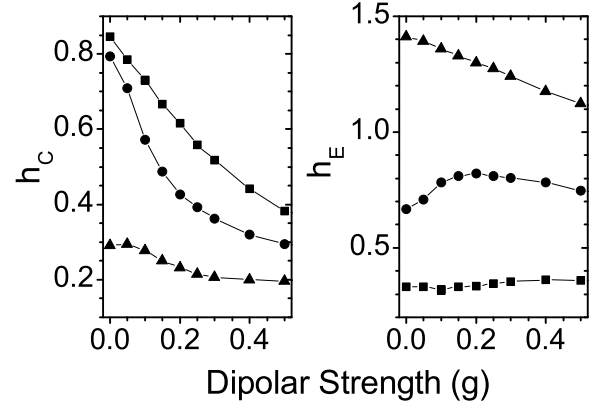}
\caption{Dependence of coercivity ($h_{C}$) and exchange bias field ($h_{E}$) on dipolar strength for a monolayer of core-shell NPs at low temperature ($t/k_C=0.01$). Curves for different values of the interface exchange are shown. Triangles: $J/k_C=1.5$, circles: $J/k_C=1.0$, and squares: $J/k_C=0.5$.  Shell anisotropy $k_S/k_C=5.0$. }
\label{f9}
\end{figure}

We discuss next the hysteresis behavior of a 2D hexagonal array of magnetic NPs with core-shell morphology.
To calculate the hysteresis loop, the spins are set initially in the saturation state along the positive $x$-axis\cite{COMMENT-2} and the field is swept from positive to negative values and back.
This choice produces a negative shift of the hysteresis loop, namely a positive exchange bias field.
The main issue we address is the dependence of the remanence, the (effective) coercivity and the exchange bias field on the strength of the DDI.
The total energy of the NP assembly is described by Eq.\ref{eq2}.
The coercivity for the shifted loops is defined as
$H_C = (1/2) |H_{C1}-H_{C2}|$,
and the exchange bias field as
$H_E = (1/2) |H_{C1}+H_{C2}|$,
where $H_{C1}$ ($H_{C2}$) is the upper (lower) branch coercivity corresponding to the backward (forward), with respect to the exchange bias field direction, magnetization reversal process.

In the MB model the microscopic details (atomic structure, defects, exchange coupling strength) are absorbed into the value of the exchange constant $J$. Thus results for different values of $J$ are considered.
The anisotropy of the AFM oxide is assumed much higher than the core anisotropy, and in the present work we take $k_S/k_C=5.0$.
This value of $k_S$ is high enough to ensure blocking of the shell magnetization $(S_{AFM})$ for applied fields in the range that the core magnetization $(S_{FM})$ exhibits hysteresis behavior.
Typical hysteresis loops for dipolar interacting NP arrays are shown in Fig.~\ref{f8}.
Similarly to the case of simple FM NPs discussed in the previous section, $M_r$ in core-shell NPs increases and $H_C$ decreases with increasing DDI strength (Fig.~\ref{f9}).
However the changes of $H_C$ due to dipolar coupling are controlled by the value of the interface exchange.
In particular, larger reduction of $H_C$ is observed in systems with larger interface exchange $J$.
Nevertheless, the fraction of $H_C$ reduction is nearly the same for all values of $J$.

\begin{figure}
\includegraphics[]{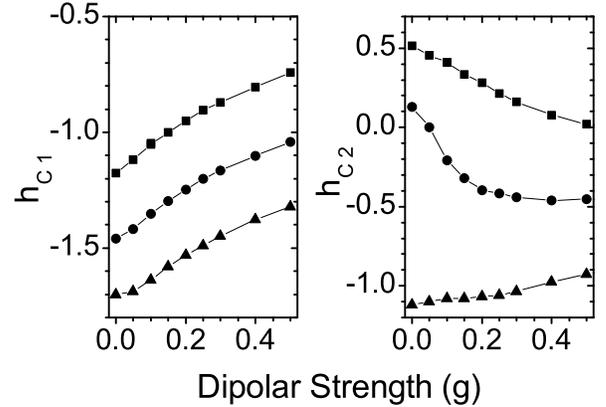}
\caption{Dependence of the upper branch coercivity ($h_{C1}$) and the lower branch coercivity ($h_{C2}$) on dipolar strength for a monolayer of core-shell NPs at low temperature ($t/k_C=0.01$). Curves for different values of the interface exchange coupling are shown. Triangles: $J/k_C=1.5$, circles: $J/k_C=1.0$, and squares: $J/k_C=0.5$.  Shell anisotropy $k_S/k_C=5.0$. }
\label{f9b}
\end{figure}

The behavior of $H_E$ with increasing dipolar strength appears more complex (Fig.~\ref{f9}).
For strong interface coupling ($J/k_C=1.5$) $H_E$ drops linearly with the dipolar strength and for weak interface coupling ($J/k_C=0.5$) it is only weakly dependent on the dipolar strength.
The most striking behavior is observed when the interface exchange is comparable to the core anisotropy ($J/k_C=1.0$).
In this case, weak DDI \emph{enhance}  $H_E$, which exhibits a maximum value around $g/k_c\simeq 0.2$.
The observed enhancement of $H_E$ due to weak DDI is contrary to what one expects on intuitive grounds. Namely, DDI that lead to symmetric backward and forward magnetization reversal processes, are shown to enhance the loop asymmetry.
Insight into the dependence of the exchange bias field on dipolar strength can be obtained from separate examination of the two branches of the hysteresis loop and the corresponding coercivity values $H_{C1}$ and $H_{C2}$.
As shown in Fig.~\ref{f9b}, $H_{C1}$ decreases (in absolute value) with dipolar strength, independently of the interface exchange value. This behavior is similar to what is observed in FM NPs (Fig.~\ref{f2}) at low temperature.
The similarity stems from the fact that in the case of backward magnetization reversal the exchange bias field  \emph{increases} the barrier height for reversal of an isolated moment. The collective reversal induced by DDI facilitates the reversal process leading to (absolute) lower coercivity values.
On the contrary, the dependence of $H_{C2}$ on dipolar strength varies significantly with the interface exchange. In the forward reversal process the exchange bias field \emph{reduces} the barrier for the reversal of an isolated moment.
When the value of the exchange is weak ($J/k_C=0.5$, Fig.~\ref{f9b}), a reduced but finite barrier exists for the forward reversal. As previously, the collective reversal, induced by DDI, reduces further the barrier height leading to smaller $H_{C2}$ values.
When the interface exchange is strong ($J/k_C=1.5$, Fig.~\ref{f9b}), the barrier for the forward reversal of an isolated moment disappears and the forward reversal becomes a "downhill" process in the energy landscape, a fact reflected in the negative value of $H_{C2}$. In this case, DDI introduce additional barriers due to their anisotropic character and the collective motion of the moments obstructs the reversal. The values of $H_{C2}$ approach zero (from negative values) as the dipolar strength increases, reflecting the increasing difficulty for forward reversal.
For intermediate interface exchange values ($J/k_C=1.0$, Fig.~\ref{f9b}) the dependence of $H_{C2}$ on the dipolar strength shows a varying behavior.
For isolated moments ($g=0$), $H_{C2}$ assumes small positive values indicating a very low barrier for forward reversal. DDI initially facilitate the forward reversal, but with increasing dipolar strength the barrier vanishes and $H_{C2}$ reaches a negative value ($g/k_C\simeq 0.2$). Further increase of the dipolar strength inhibits the forward reversal.
The observed dependence of $H_E$ on DDI strength shown in Fig.~\ref{f9} is the net effect from the variation of the upper and lower branch coercivities with dipolar strength.
Experimentally, enhancement of $H_E$ due to DDI has been observed\cite{HBI03} in stripes of Co/CoO NPs where the quasi-one-dimensional morphology of the array enhances the effect of DDI.

\subsection{Transverse susceptibility of FM nanoparticles}

\begin{figure}
\includegraphics[]{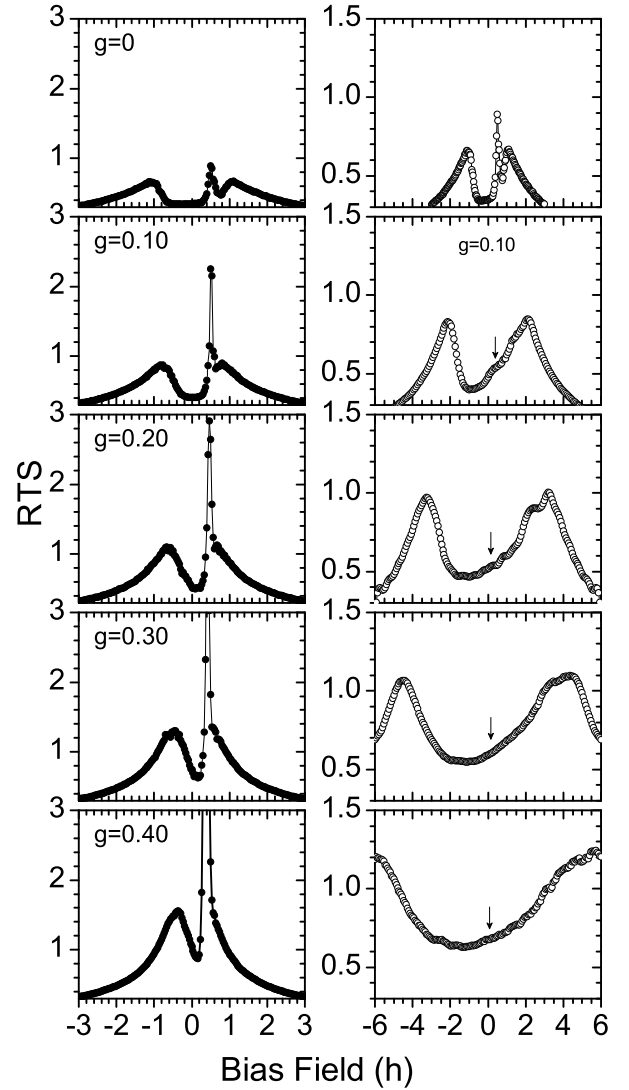}
\caption{Field dependence of reversible transverse susceptibility of a monolayer of FM nanoparticles at low-temperature ($t/k=0.1$). The bias field is swept from negative to positive values. Closed circles: in-plane field. Open circles: out-of-plane field. The arrows in the out-of-plane data indicate the position of the coercive field.}
\label{f10}
\end{figure}

\begin{figure}
\includegraphics[]{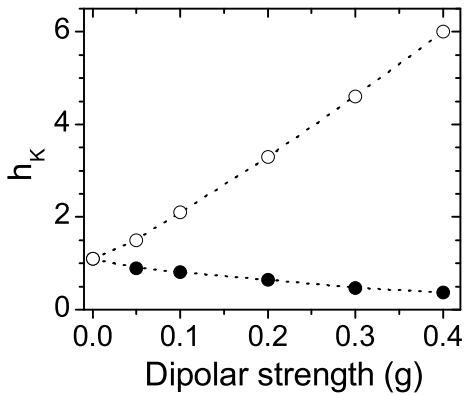}
\caption{Dependence of anisotropy field on dipolar strength at low temperature ($t=0.05$) for a monolayer of FM nanoparticles. Closed circles: in-plane field. Open circles: out-of-plane field.}
\label{f11}
\end{figure}

Measurements of the field dependent RTS have been long ago predicted theoretically\cite{AHA57} to reveal direct information on the magnetic anisotropy of a NP assembly. A typical RTS curve of a NP assembly with random anisotropy, obtained at low temperature by sweeping the bias (dc) field in one direction, exhibits three characteristic peaks, one at the coercive field and two at positions corresponding to the anisotropy field $(\pm H_K)$. The latter is related to the single-particle anisotropy as $H_K=2K_1/M_s$. Thus measurement of RTS should, in principle, be a direct method to measure $K_1$. However, wide particle size distributions\cite{HOA93}, thermal fluctuations\cite{CHA94,YAN95,KEC06} and most importantly, dipolar interaction effects\cite{SOL97,CHA94,YAN95,KEC06} modify the position and shape of the peaks making the determination of the anisotropy strength uncertain.
We discuss here the evolution of the field dependent RTS curves under increasing values of the dipolar strength, or equivalently decreasing values of the interparticle distance in SAA of magnetic NPs.
We show in Fig.~\ref{f10} the in-plane $(\chi_T^{\parallel})$ and out-of-plane $(\chi_T^{\perp})$ RTS curves at low temperature $(t/k = 0.05)$ for increasing dipolar strength. The non-interacting sample shows clearly the theoretically predicted three peaks located at the coercive $(h_c)$ and anisotropy $(\pm h_K)$ fields.  Downshifted values of $h_c$ and $h_K$ relative to the zero-temperature values ($h_c=0.98$ and $h_K=2$) are due to thermal fluctuations effects.
The most important effects of dipolar interaction on the RTS curves, shown in Fig.~\ref{f10} and \ref{f11} are :
(i) the suppression of the $h_c$ peak of $\chi_T^{\perp}$,
(ii) the location of the $h_K$ peak of $\chi_T^{\perp}$ at higher fields than the corresponding peaks of $\chi_T^{\parallel}$,
(iii) the downshift (upshift) of the $h_K$ peak of $\chi_T^{\parallel}$ ($\chi_T^{\perp}$) with increasing dipolar strength,
(iv) the slower saturation with bias field of $\chi_T^{\perp}$  relative to $\chi_T^{\parallel}$.
In recent RTS measurement in Fe NP arrays\cite{POD03}, observations similar to our points (ii) and (iv) were made. However, the coercivity peak could not be resolved probably due to not sufficient lowering of the temperature or due to the presence of NP size distribution.
Nevertheless, the agreement of our results in points (ii,iv) above, constitute sufficient evidence that DDI were responsible for the observed experimental trends in these measurements.
A physical interpretation of the observed opposite trends of the in-plane and out-of-plane $h_K$ peaks with increasing dipolar strength (Fig.~\ref{f11}) relies on the development of an easy-plane anisotropy induced by DDI.
For an in-plane bias field, the interaction-induced anisotropy reduces the barrier for an irreversible switching of the moments leading to a reduction of the anisotropy field, while in the out-of-plane geometry DDI inhibit the irreversible switching along the $z$-axis by developing an easy-plane normal to this axis, thus increasing the anisotropy field.
Finally, the linear dependence of the anisotropy peaks on the dipolar coupling, or equivalently, on the inverse cube of the interparticle spacing ($H_K \sim 1/d^3$), shown in Fig.~\ref{f11} for $g/k \leq 0.4$, could be used to perform an extrapolation procedure on measurements taken at different interparticle separations in order to extract the value of single-particle anisotropy $(K_1)$.

\subsection{Tunneling magnetoresistance in FM nanoparticle arrays}

\begin{figure}
\includegraphics[]{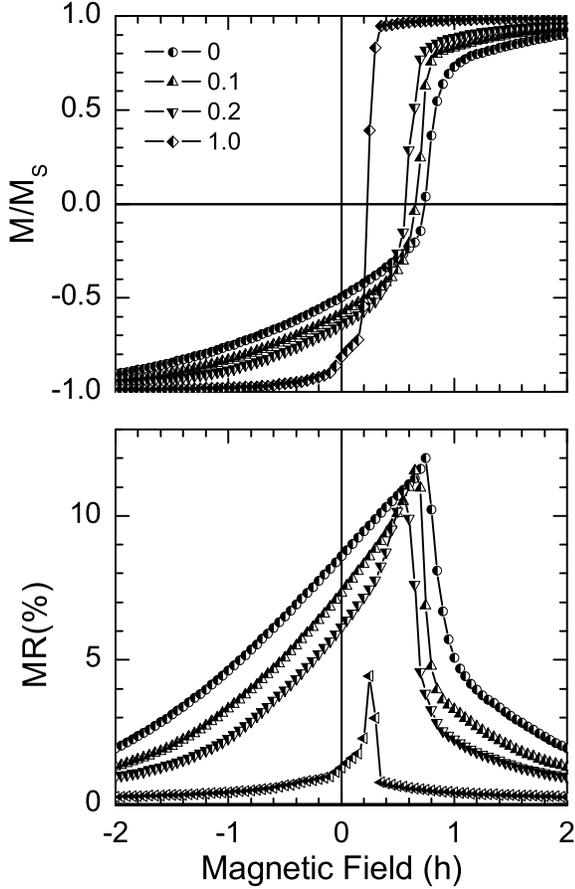}
\caption{Field-dependence of the magnetization (upper panel) and the corresponding tunneling magnetoresistance (lower panel) of a monolayer of FM nanoparticles, at low temperature ($t/k=0.01$). The curves correspond to different dipolar strengths. Circles: $g=0$, up-triangles: $g/k=0.1$, down-triangles: $g/k=0.2$, and diamonds: $g/k=1$. The bias field lies in-plane and is swept from negative to positive values. }
\label{f12}
\end{figure}

Tunneling magnetoresistance refers to a large decrease of a sample's resistivity under application of a bias magnetic field, observed when charge carriers transmit through two FM regions separated by a non-magnetic insulating barrier. The basic mechanism underlying the effect is  spin-dependent scattering of the carriers. The first FM region acts as a polarizer for the electron spin while the second region causes scattering whose strength is proportional to the misalignment of the magnetization relative to the first region. In the case considered here, the magnetic NPs are the relevant FM regions and the surfactant layer separating them is the insulating barrier. Thus, in principle, spin-dependent transport measurements should reflect the underlying micromagnetic structure of the NP assembly.
In Fig.~\ref{f12} we plot the field dependent TMR of a monolayer of dipolar coupled NPs and compare it with the corresponding branch of the magnetization hysteresis loop.
The sharp peak of TMR occurs very  close to the coercive field, because the spin disorder  in the array is maximized at his field. The effects of DDI can be observed in the TMR curves. A downshift of the TMR peak position with increasing dipolar strength is observed following the reduction of the $H_c$ values. The value of TMR at the remanent state decreases with interactions, reflecting the increasing alignment of the magnetic moments demonstrated also by the increased values of the $M_r$ (see Fig.~\ref{f2}). Finally, the TMR sensitivity, namely the slope of the field dependent TMR, increases with increasing interaction strength, due to a collective reversal of the moments during which the degree of alignment is higher for stronger interactions.

\begin{figure}
\includegraphics[]{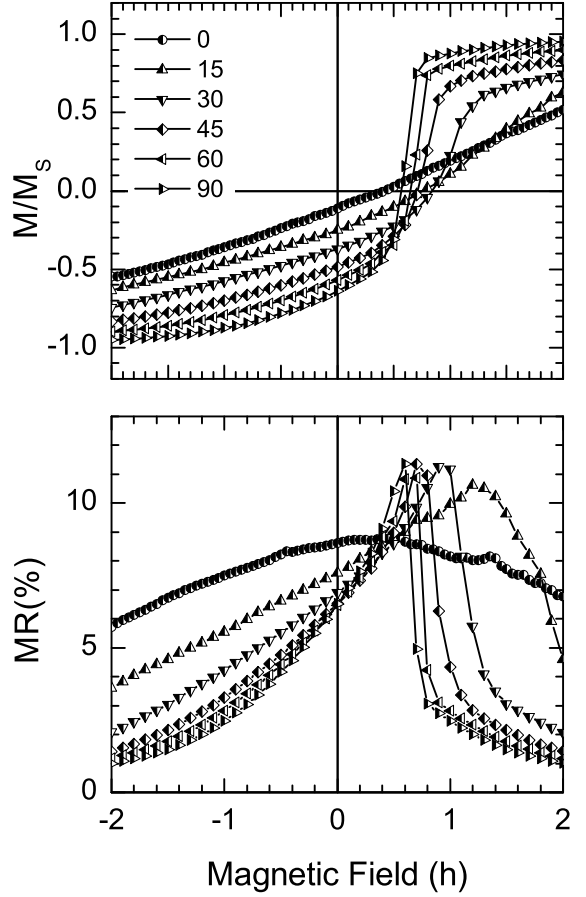}
\caption{Field-dependence of the magnetization (upper panel) and the corresponding tunneling magnetoresistance (lower panel) of a monolayer of dipolar interacting ($g/k=0.2$) FM nanoparticles, at low temperature ($t/k=0.01$). The curves correspond to different directions of the bias field relative to the z-axis ($\theta=0^{\circ}-90^{\circ}$). The field is swept from negative to positive values.  }
\label{f13}
\end{figure}

The easy-pale anisotropy induced by DDI in a 2D-array of NPs is expected to produce a strong dependence of the TMR values on the direction of the bias field. Indeed the TMR curves shown in Fig.~\ref{f13}  vary substantially with the azimuth angle $(\theta)$. The TMR sensitivity decreases as the field approaches the $z$-axis, reflecting the slow saturation of the magnetization for an out-of-plane bias field.
A striking feature occurring for $\theta\simeq15^{\circ}$ in Fig.~\ref{f13} is that the peak of TMR does \emph{not} occur at the coercive field but at a higher field.
We state that the peak of TMR occurs at the critical field $(H_o)$, namely the field for an irreversible switch of the magnetization,\cite{CHKZ64} rather than at the coercive field. This has been verified by simulations in purely dipolar arrays\cite{KEC05} or non interacting arrays with aligned easy axes. In both cases the critical field can be obtained analytically and it was found that the peak of TMR occurs exactly at this field. When DDI are absent, the random anisotropy leads to $H_c\approx 0.96 H_o$\cite{CHKZ64}  and the TMR peak occurs very close to the coercive field.  However, as DDI increase, they induce coherent rotation of the moments and a dominant easy-plane anisotropy.
In the case of an easy-plane anisotropy the difference between the $H_o$ and $H_c$ is maximum for $\theta\approx 15^{\circ}$\cite{CHKZ64}, which explains why the maximum deviation between the field corresponding to the peak of TMR and the coercive field occurs at this angle.

\section{FINAL REMARKS}

We have given an overview of our simulation studies on the effects of dipolar interactions on the magnetization, the short-range correlation function, the transverse susceptibility and the conductivity of nanoparticle assemblies forming hexagonal arrays. We discussed the modifications introduced to the field and temperature dependence of the above quantities due to the dipolar coupling between nanoparticles.
We have demonstrated that dipolar interactions in thin layers formed by self-assembled NPs  induce an easy-plane anisotropy that is responsible for an anisotropic magnetic behavior with respect to the bias field direction. The anisotropic behavior is revealed in different physical quantities used to magnetically characterize nanoparticle arrays, such as the transverse susceptibility and the tunneling magnetoresistance and in the values of characteristic magnetic parameters, such as the remanence, the coercivity and the blocking temperature.
The most important dipolar interaction effects predicted by our numerical studies and observed in experiments are summarized as follows :
(i) The increase of remanence with particle density for submonolayer coverage, observed in self-assembled Fe NPs.\cite{FAR05}
(ii) The enhancement (suppression) of the in-plane (out-of-plane) remanence at low temperatures, in agreement with magnetization measurements on self-assembled Co NPs.\cite{PIL01}
(iii) The suppression of the coercivity at low temperatures with decreasing interparticle spacing, observed in self-assembled Fe NPs.\cite{FAR05}

(iv) The increase of the apparent blocking temperature with decreasing interparticle separation $(T_b \sim 1/d^3)$, in agreement with ZFC magnetization measurements on self-assembled $\epsilon$-Co NPs\cite{ZHA03}
(v) The existence of two distinct blocking temperatures for the in-plane and out-of-plane geometries, in agreement with ZFC magnetization  and transverse susceptibility measurements on self-assembled Fe NPs.\cite{POD03}
(vi) The persistence of short-range FM correlations at temperatures above the blocking existence; x-rays measurements on Co NPs\cite{KOR05} showed AFM correlations at high temperatures, while SANS measurements on Fe NPs did not provide evidence of AFM correlations.\cite{IJI05} More refined experiments are required in this direction.
(vii) The possibility to enhance the exchange bias field with decreasing interparticle separation at low temperatures in qualitative agreement with the enhanced exchange bias field found in oxidized Co NPs forming chain-like structures.\cite{HBI03}
(viii) The decrease (increase) of the anisotropy field, obtained from transverse susceptibility measurements with an in-plane (out-of-plane) bias field, with decreasing interparticle separation, in agreement with measurements on self-assembled Fe NP samples.\cite{POD03}

Certain results of our simulations that are yet to be verified by experiments include :
(i) The transition from a 2D to 3D magnetization reversal mechanism with increasing number of monolayers that is responsible for a reduction of the coercivity with number of stacked monolayers. A layer-by-layer measurement of the magnetic properties would address this point.
(ii) The suppression of the TMR values and the \emph{increase} of TMR sensitivity in dense arrays. Verification of this result requires comparison between TMR measurements in dilute samples and self-assembled arrays, and finally
(iii) The location of the TMR peak \emph{not} at the coercive field, for a tilted bias field in  dense assemblies
New developments in self-assembly techniques\cite{LID04} will make feasible further investigation of the interplay between geometry and interaction effects in magnetic NP arrays, while multiscale computational schemes\cite{GAR06} constitute a promising tool to achieve more detailed understanding and control of their properties.

\begin{acknowledgments}
The authors acknowledge S. A. Majetich and G. Markovich for discussions on the chemical preparation of self-assembled samples and their structural and magnetic characterization, and H. Srikanth for his comments on the transverse susceptibility measurements. Work supported by EC project NANOSPIN (Contract No NMP4-CT-2004-013545) and the IMS Center of Excellence on Nanostructured Materials (Project No. 962)
\end{acknowledgments}


\begin{thebibliography}{96}
\bibitem[{COM({\natexlab{a}})}]{COMMENT-1}
\bibinfo{note}{The MB model predicts that the coercivity is suppressed due the
  inteface exchange bias field, namely $H_c(J=0) > H_c(J\neq 0)$. This result
  is in contrast to the experimental situation, where the coercivity is
  enhanced under field-cooling. To reproduce this trend one should extend the
  MB model to include the interface moment due to uncompensated spins.
  \cite{EFT05}}.
\bibitem[{COM({\natexlab{b}})}]{COMMENT-2}
\bibinfo{note}{The alignement of spins along the $x$-axis corresponds to the
  final state of the system when a cooling process under a strong field
  ($h_{FC} \gg k_S, k_C$) along the $+x$-axis is performed prior to measuring
  the hysteresis loop. Under this assumption the maximum exchange bias effect
  is observed.}
\end{thebibliography}
\end{document}